\documentclass{jfm}

\usepackage{graphicx}
\usepackage{newtxtext}
\usepackage{newtxmath}
\usepackage{natbib}
\usepackage{hyperref}

\begin{document}

\title{Circular hydraulic jumps: where does surface tension matter?}

\author{A.~Duchesne \aff{1}
 \corresp{\email{alexis.duchesne@univ-lille.fr}}  
\and L.~Limat \aff{2}
}

\affiliation{
\aff{1} Univ. Lille, CNRS, Centrale Lille, Univ. Polytechnique Hauts-de-France, UMR 8520 - IEMN, F-59000 Lille, France
\aff{2} Université de Paris, CNRS, Laboratoire Mati\`ere et Syst\`emes Complexes (MSC), UMR 7057 - B\^atiment Condorcet, 10 rue Alice Domon et L\'eonie Duquet, 75013 Paris, France}

\maketitle

\begin{abstract}
Recently, an unusual scaling law has been observed in circular hydraulic jumps and has been attributed to a supposed missing term in the local energy balance of the flow [\cite{bhagat_2018}].  In this paper, we show that - though the experimental observation is valuable and interesting - this interpretation is presumably not the good one. When transposed to the case of a axial sheet formed by two impinging liquid jets, the assumed principle leads in fact to a velocity distribution in contradiction with the present knowledge for this kind of flows. We show here how to correct this approach by keeping consistency with surface tension thermodynamics: for Savart-Taylor sheets, when adequately corrected, we recover the well known $1/r$ liquid thickness with a constant and uniform velocity dictated by Bernoulli's principle.

In the case of circular hydraulic jumps, we propose here a simple approach based on Watson description of the flow in the central region [\cite{Watson_JFM}], combined with appropriate boundary conditions on the formed circular front. Depending on the specific condition, we find in turn the new scaling by \cite{bhagat_2018} and the more conventional scaling law found long ago by \cite{Bohr_1993}. We clarify here a few situations in which one should hold rather than the other, hoping to reconcile Bhagat et al. observations with the present knowledge of circular hydraulic jump modeling. However, the question of a possible critical Froude number imposed at the jump exit and dictating logarithmic corrections to scaling remains an opened and unsolved question. 
\end{abstract}

\section{Introduction}
\begin{figure}
\begin{centering}
\includegraphics[width=0.6 \columnwidth]{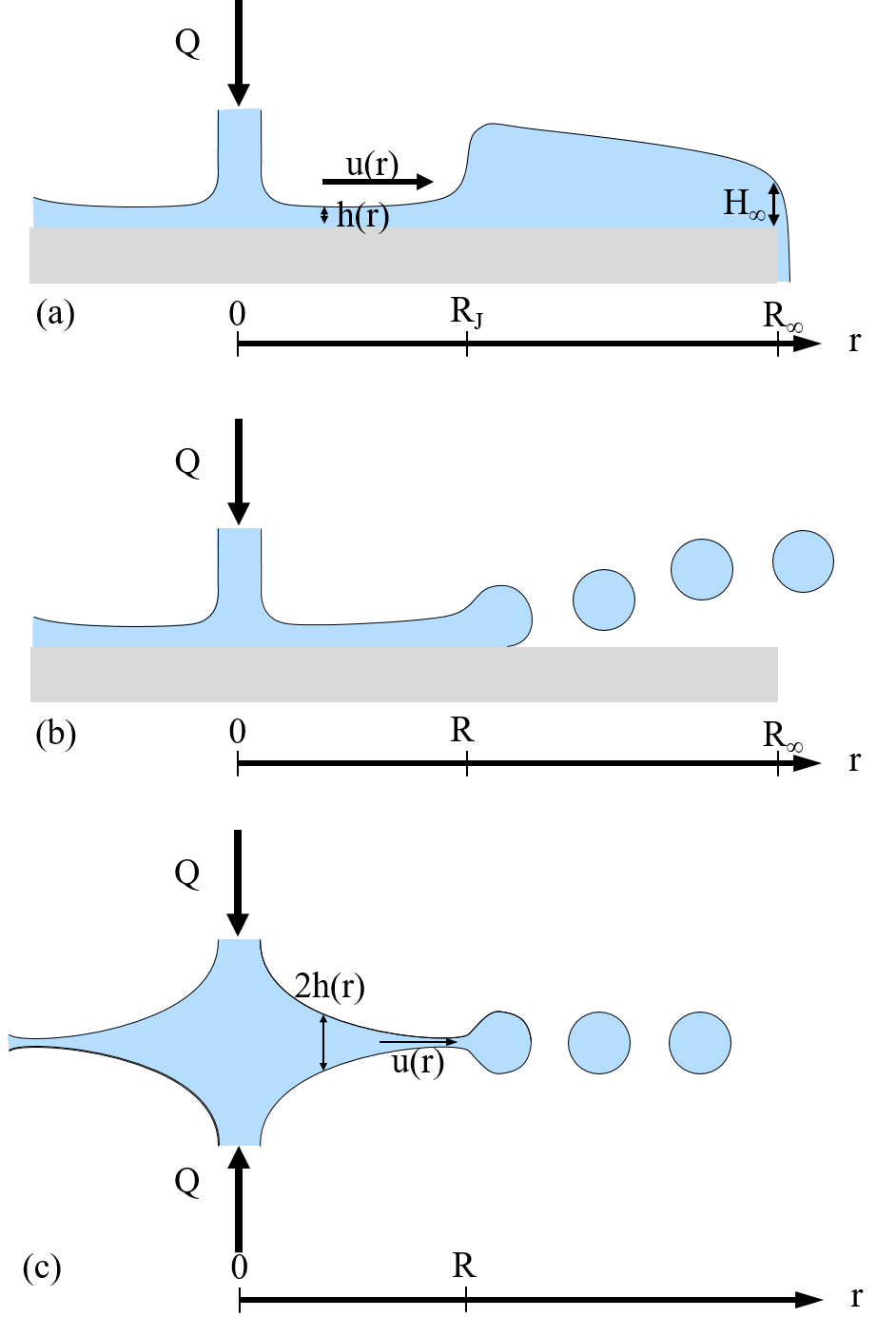}
\caption{Three axisymmetric film flows are discussed in the present article. (a) The classical circular hydraulic jump formed by a jet impacting a solid disk at its center, (b) atomization ring formed by a jet impacting a dry surface possibly superhydrophobic, (c) liquid sheet formed by impact of two liquid jets of opposite direction.  }
\label {fig1}
\end{centering}
\end{figure}

Stationary axisymmetrical liquid structures formed by jet impacts, have motivated an enormous amount of literature. Three examples that will be important here are sketched on Fig. 1. First of all, the well-known circular hydraulic jump [\cite{Rayleigh_1914, Tani_48, Watson_JFM, Craik_81, Bohr_1993, Bush_2003,  duchesne_2014, mohajer_2015, salah_2018, bhagat_2018,  wang_2019,  wang_2021}], sketched on Fig.\ref{fig1}-a, with a well developed liquid film extending all around. Its equivalent on a "dry" surface, possibly superhydrophobic [\cite{jameson_2010, button_2010, Maynes_2011}], the "rim atomization" is sketched on Fig.\ref{fig1}-b. Finally the well-known radial liquid sheet [\cite{savart_1833_2, huang_1970,  clanet_2002, villermaux_2013}], formed either by impinging two opposite symmetrical liquid jets, having the same central axis or by impinging a liquid jet on a solid surface with a diameter similar to the jet diameter is depicted on Fig.\ref{fig1}-c.

These three geometries are of course linked together by the same general equation for the energy balance. In this article, we will therefore show that apparent paradoxes raised by the modeling of the surface tension on the circular hydraulic jump by \cite{bhagat_2018} may be solved or at least clarified by considering the geometry depicted in Fig.\ref{fig1}-c.

The selection of jump radius $R_J$ in the circular hydraulic jump case (Fig.\ref{fig1}-a) has motivated many studies. The two most well known approaches are the one from Watson and Bush [\cite{Watson_JFM, Bush_2003}], in which the height of the outer film remains a control parameter, and the one from \cite{Bohr_1993}, rather devised when a liquid film extends all around at large distance, and inspired from boundary layer theories. As well known, this second approach leads to a scaling law dependence of $R_J$ upon flow rate $Q$ and the physical parameters ($\nu$  the kinematic viscosity of the fluid, $g$ gravity), that reads:

\begin{equation}
R_J\sim Q^\frac58 \nu^{-\frac38}g^{-\frac18}.
\label{eq.1}
\end{equation}

Later, \cite{duchesne_2014}  emphasized the importance of logarithmic corrections to scaling, due to viscous dissipation in the outer film, yet observed numerically by Bohr, and also showed that the prefactor was experimentally linked to the value of the Froude number at the exit of the jump, that seemed to be locked to a critical value. This phenomenon was recovered by \cite{mohajer_2015} and by \cite{argentina_2017} with a non-linear modeling of film flow equations including the first finite slope terms. 

Very recently, an attempt of revision of this picture has been published by \cite{bhagat_2018}, who performed new experiments, and reported the observation of a different scaling in which surface tension $\gamma$ was involved, but not gravity:

\begin{equation}
R_J\sim Q^\frac34  \rho^\frac14 \nu^{-\frac14}\gamma^{-\frac14},
\label{eq.2}
\end{equation}

in which $\rho$ is the liquid density. To rationalize this finding, these authors claimed that most available approaches of surface tension influence lead to only small corrections [\cite{Bush_2003}] and that the description of the circular hydraulic jump had thus to be completely reconsidered. They introduced an energy balance, between two radii $r$ and $r+\delta r$, that reads:

\begin{equation}
\left[\rho \frac{\bar u^2}2 \bar u rh \right]^{r+\delta r}_r=\left[\gamma r\bar u  \right]^{r+\delta r}_r-\left[p \bar u h \right]^{r+\delta r}_r-\left[\rho g \frac{h^2}2 r \bar u  \right]^{r+\delta r}_r-r \tau_W\bar u \delta r,
\label{eq.3}
\end{equation}

with the notation $\left[A \right]^{r+\delta r}= A(r+\delta r)-A(r)$, and in which $\bar u$ designates the flux-average radial velocity, $r$ the distance to the axis, $h(r)$ the thickness of the liquid layer, $p(r)$ the pressure at $z=0$ and $\tau_W$ the wall shear stress. The last term on the right designates the viscous dissipation by friction on the substrate, while the first one is an additional term compared to previous approaches, that is presumed to be “at the origin” of the new scaling (\ref{eq.2}). This conjecture has been contested [\cite{duchesne_2019, bohr_2021}] (see also\cite{bhagat_2020} answer), and it is also known that a scaling like (\ref{eq.2}) can also appear without such an assumption, as shown for instance by\cite{button_2010} for liquid bells formed below a ceiling.  

We find here useful to have a look on what would happen in the simplified geometry of Fig.\ref{fig1}-c, when applying this principle. As we shall show in section \ref{sec.2}, this modeling leads to a velocity distribution in complete contradiction with the present knowledge of liquid sheets (and with Bernoulli's principle), which suggests that Bhagat et al argument is flawed. In fact the obtained flow field is not new, and has been proposed in the past by \cite{Bouasse_1923} who attributed the calculation to Hagen (see \cite{hagen_1849}). It will be instructive here to remind the argument followed by Hagen and Bouasse, in section \ref{sec.3}, in a Lagrangian frame, analyzing a circular expanding piece of film. We will then show, in the same section, how one can correct the argument to get the more classical and now admitted result deduced from Bernoulli's principle of a uniform radial velocity around the impact point, and how, missing some terms in the balance, one can get the flawed result of Bouasse and Hagen. Finally, coming back to a Eulerian description, we will explain how these considerations impact the principle proposed in eq.(\ref{eq.3}). We will show that an extra term exactly equal and opposite to the capillary contribution  should cancel this one, in a way consistent with classical thermodynamics, leading to the expression usually written from the balance of momentum.  

This does not mean, however, that Bhagat  et al. scaling discovery is of none interest. In section \ref{sec.4} and  \ref{sec.5}, we will try to precise to which capillary structures – different from the stationary hydraulic jump observed by Bohr – it could apply, and a possible way to justify its occurrence.

\section {A look to a simple situation: the axisymmetrical liquid sheet.}\label{sec.2}

Let us try to apply the principle suggested in eq.(\ref{eq.3}) to the case suggested on Fig.\ref{fig1}-c, i.e. to a axisymmetrical sheet formed by the coaxial impact of two jets in a situation of negligible gravity. The viscous shear on the substrate having disappeared, eq.(\ref{eq.3}) reduces to a very simple balance that reads:

\begin{equation}
\left[ \rho rh \frac{ u^3}2 \right]^{r+\delta r}_r=\left[ \gamma r u  \right]^{r+\delta r}_r,
\end{equation}

where the horizontal velocity $u$ has no dependence upon the transverse direction, and coincides with any of its average values.  This implies that the following quantity is constant all over the sheet:

\begin{equation}
\rho rh \frac{ u^3}2-\gamma r u=\mathrm{Cte}.
\end{equation}

Combined with the mass balance $Q=2\pi rhu$, this leads to the following expression for $u$:
\begin{equation}
u=2 \pi \frac\gamma{\rho Q}r+\sqrt{u_0^2-4 \pi \frac\gamma{\rho Q}r_0 u_0+ 4 \pi^2 \frac{\gamma^2}{\rho^2 Q^2}r^2},
\end{equation}

in which $r_0$ designates the jet radius at impact and $u_0$ the asymptotic value for $u$, reached when $r=r_0$, which satisfies the equality $Q=\pi r_0^2 u_0$ in a quasi-elastic shock approximation [\cite{villermaux_2013}].  In the limit of large jet velocity, i.e. $u_0^2 \gg 2\gamma⁄(\rho r_0 )$, this expression reduces to the slowly varying upon $r$ approximate:

\begin{equation}
u\approx u_0+ 2 \pi \frac\gamma{\rho Q}(r-r_0),
\label{eq.7}
\end{equation}

which is known to be false, as it has been checked experimentally that the velocity is constant all over the sheet, recovering the Bernoulli's principle (see in particular Fig. 3 in [\cite{villermaux_2013}]). It is however amazing to remind that a similar expression is proposed by \cite{Bouasse_1923} who attributed this result to \cite{hagen_1849}, but with a slight sign change, that is in fact due to a mistake on his own:

\begin{equation}
u\approx u_0-2 \pi \frac\gamma{\rho Q}(r-r_0).
\label{eq.8}
\end{equation}

Though obtained erroneously, this expression is very seductive and Bouasse used it to calculate the radius of the liquid sheet $R_{LS}$ assuming that the sheet border should stay at the place in which $u$ vanishes which leads to $R_{LS}=(\rho Q u_0)⁄2\pi \gamma  (=\rho r_0^2 u_0^2)⁄2\gamma$ . Surprisingly this result coincides with the right one that is in fact obtained, now, by assuming a constant velocity, dictated by Bernoulli's principle, and the balance of momentum at the sheet perimeter, i.e. $\rho hu^2=\gamma$ [\cite{villermaux_2013}]. But on the other hand, we would like to stress out that the radial velocity is uniform in the sheet of Fig.\ref{fig1}-c, which means that the principle proposed in eq.(\ref{eq.3}), and therefore the basis of the theory developed by \cite{bhagat_2018} is flawed.

\section {Reconsidering Hagen argument, and its implications for hydraulic jump.}\label{sec.3}

We try to understand the fault underlying Bouasse and Hagen principle. Their line of thought is easier to explain considering a Lagrangian frame, and more precisely the balance of energy on a annular piece of fluid, convected by the radial flow, and it is in fact the method proposed by Bouasse himself in his treatise of fluid mechanics [\cite{Bouasse_1923}].  

\begin{figure}
\begin{centering}
\includegraphics[width=0.6 \columnwidth]{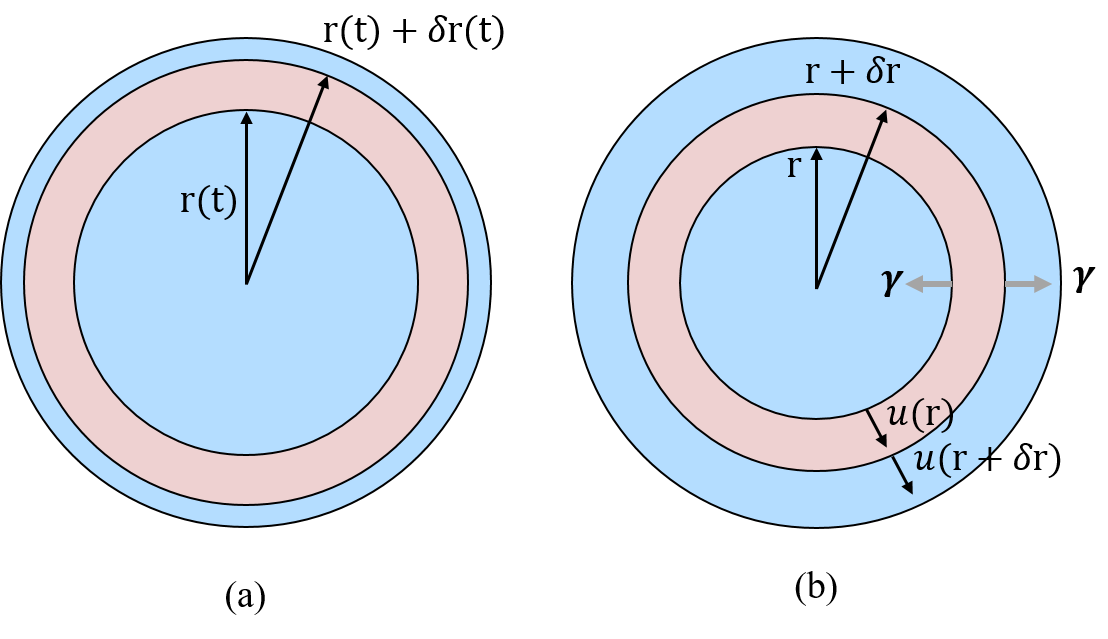}
\caption { Lagrangien (a) and Eulerian (b) frame for discussing energy balance in a annular portion of a liquid film used in the text.}
\label {fig2}
\end{centering}
\end{figure}

Let us consider a piece of annular piece of film as on Fig.\ref{fig2}-a, convected and distorted by the flow. Mass conservation implies that, at any time  $hr\delta r=\mathrm{Cte}$, while the balance of energy for the whole annulus reads, in the limit of $\delta r$ small enough to satisfies the condition $\delta r \frac {\partial u} {\partial r} \ll u$   of a slowly varying velocity field :

\begin{equation}
\frac {\partial} {\partial t} \left[2 \pi \left(\frac 12 \rho u^2rh \delta r+\gamma r \delta r \right) \right]\approx 2 \pi \gamma \delta r u.
\label{eq.9}
\end{equation}

The first term in the left hand side of this equation stands for kinetic energy, and the second for the surface energy enclosed between $r$ and $r+\delta r$. The right hand term comes from the work of surface forces, and does not vanish. Indeed, the same surface tension force is pulling on a different arc length, as the external boundary has a  larger perimeter than the other (note that this is the intuitive argument underlying Bhagat's analysis). Still in the limit of a slowly varying velocity field at the scale $\delta r$, after noting that $\frac {\partial } {\partial t}=u \frac {\partial } {\partial r}$,  eq.(\ref{eq.9}) reads:

\begin{equation}
rh \delta r u \frac {\partial } {\partial r}\left(\frac{\rho u^2}2\right)+\gamma u \delta r\approx \gamma u \delta r.
\label{eq.10}
\end{equation}

In fact, the two surface tension terms are canceling each other, which means that the work provided to the annulus by surface tension of the outer interfaces is completely transformed into the surface energy stored at the free surface of the annulus, in agreement with simple thermodynamic considerations. As a result, the fluid velocity is unaffected by surface tension balance and remains constant as one would deduce from a more classical argument in terms of Bernoulli's principle, i.e. $u(r)$ is in fact independent of $r$:

\begin{equation}
u(r)=u_0= \mathrm{Cte}.
\end{equation}

Note here that skipping from eq.(\ref{eq.9}) to eq.(\ref{eq.10}) is not completely trivial as there is an extra $\gamma$ term remaining, but this one vanishes for the constant and uniform $u_0$ solution. 

To reconnect with Bouasse, instead of this, if one would forget the internal surface energy contribution in the left hand member of eq.(\ref{eq.10}), one would get the following equation for $u$, that reduces to:

\begin{equation}
rh \delta r \frac {\partial } {\partial r}\left(\frac{\rho u^2}2\right)\approx \gamma  \delta r.
\end{equation}

After simplifying $\delta r $, and using the fact that $Q= 2\pi r u h$, this equation leads to: 

\begin{equation}
 \frac {\partial  u} {\partial r}\approx 2 \pi \frac \gamma {\rho Q},
\end{equation}

which leads finally to eq.(\ref{eq.7}). Alternatively, eq.(\ref{eq.8}) is obtained when one forgets the work provided to the annulus by the outer parts of the liquid sheet, i.e. by neglecting the right hand member of eq.(\ref{eq.10}), following the intuitive but erroneous idea of \cite{hagen_1849} that surface tension could slow down the flow. Historically Bouasse followed the first argument, but committed a sign mistake, obtaining eq.(\ref{eq.8}), that was physically more natural, considering presumably what Hagen said long ago. 

To summarize, a correct treatment of the expansion of liquid annula in the flow leads to the classical result of a uniform velocity, while the approximates defended by Hagen and Bouasse would follow from neglecting a part of capillary terms. We believe that a similar problem is involved in eq.(\ref{eq.3}). If we consider now a Eulerian description of the flow, as suggested on Fig.\ref{fig2}-b, the balance of energy should rather read:

\begin{equation}
\left[\rho \frac{\bar u^2}2 \bar u rh +\gamma r\bar u \right]^{r+\delta r}_r=\left[\gamma r\bar u  \right]^{r+\delta r}_r-\left[p \bar u h \right]^{r+\delta r}_r-\left[\rho g \frac{h^2}2 r \bar u  \right]^{r+\delta r}_r-r \tau_W\bar u \delta r,
\end{equation}

in which we have added in the left hand side the surface energy convected by the film. It is true that one can consider a capillary force, as in Bhagat et al, in the right hand member, but in this case, one should not miss the flux of surface crossing the two circles displayed on Fig.\ref{fig2}-b in the left hand side of the equation. And just as what happens in a Lagrangian frame, the physics being the same in both frame, the capillary effects should exactly compensate each other in this equation, that should then reduce to the more conventional form:

\begin{equation}
\left[\rho \frac{\bar u^2}2 \bar u rh  \right]^{r+\delta r}_r=-\left[p \bar u h \right]^{r+\delta r}_r-\left[\rho g \frac{h^2}2 r \bar u  \right]^{r+\delta r}_r-r \tau_W\bar u \delta r,
\end{equation}

that apart some coefficients that will depend on the detailed structure of the flow profile is consistent with what people are used to write starting rather from the balance of momentum [\cite{Bohr_1993}]. Therefore, we do not consider, in the interpretation of  eq.(\ref{eq.2}), that one should add a new capillary force distributed all over space as proposed by \cite{bhagat_2018}. To our opinion, this would imply to redo the initial mistake of Hagen and Bouasse. Just as for the calculation of the size of radial liquid sheets, the solution should rather lies inside the boundary conditions written at the circle which radius is under question. We are now developing more this idea.

\section{Alternative explanation of unusual scaling: the boundary condition at the "jump" radius. Comparison with atomization rings. }\label{sec.4}

To interpret the occurrence of  \cite{bhagat_2018} scaling, we propose an alternative approach. We just treat the two ideal situations of Fig.\ref{fig1}-a and Fig.\ref{fig1}-b with the same method, and see what happens. We will then see that the situation obtained in Fig.\ref{fig1}-b may be compared to the one suggested by \cite{bhagat_2018}.

To simplify the analysis, the “internal” flow for $r_0< r< R_J$ is assimilated to the one discussed long ago by \cite{Watson_JFM}, in which fluid inertia is progressively dissipated by viscous friction, i. e. for $r<R_J$:

 \begin{equation}
 \label{e.16}
u(r,z)=\frac{27c^3}{8\pi^4}\frac{Q^2}{\nu(r^3+l^3)} f\left(\frac z h\right),
\end{equation}

in which $c\approx 1.402$, $l=0.567r_0R$ (with $R$ the Reynolds number of the jet) and $f$ is the function: $f(\eta)=\sqrt{3}+1-\frac{2\sqrt{3}}{1+cn(3^\frac{1}{4}c(1-\eta))}$. Mass conservation implies that the film thickness and the flux of momentum are given by:

   \begin{equation}
\label{e.19}
\rho h <u^2>= \frac{27\sqrt{3}c^3}{16\pi^6}\frac{\rho Q^3}{R_J \nu(R_J^3+l^3)}, 
 \end {equation}

where $<u^2>=\int_0^h u^2 \mathrm{d}z$. In the case of  Fig.\ref{fig1}-a, this flow must be matched for $r>R_J$ to a film flow under the action of gravity, that, according to lubrication [\cite{duchesne_2014}], has a thickness distribution $H(r)$ given by:

   \begin{equation}
\label{e.20}
H(r)^4=H_\infty^4+\frac{6}{\pi}\frac{\nu Q}{g} ln \left(\frac {R_\infty} r\right),
 \end {equation}

where $R_\infty$ designates the outer radius of the substrate, where the thickness $H$ reaches a value called $H_\infty$  that will depend on the specific geometrical conditions of the flow there (see Fig.\ref{fig1}-a for the graphical definition). At $r=R_J$, one has to write some matching condition, that is consistent with the approximations made on each side of $r=R_J$, and stands for a shock [\cite{belanger_1841, Rayleigh_1914}]. If we assume $h\ll H$ and neglect the surface tension at the shock (i.e. for circular hydraulic jumps large enough such as the ones considered by  \cite{bhagat_2018}), this shock condition reads:

   \begin{equation}
\label{e.21}
\rho h (r) <u(R_J)^2> \approx \rho g H(R_J)^2,
 \end {equation}

In the limit of negligible values for  $H_\infty$ and $r_0$, compared to the other scales, it is easy to check that these equations lead to the following scaling law for $R_J$:

\begin{equation}
\label{e.22}
R_J ln \left(\frac {R_\infty} {R_J}\right)^\frac 1 8= \frac{(3c)^\frac 3 4}{2^\frac  9 8 \pi^\frac 11 8} \frac{Q^{\frac 5 8}}{\nu^{\frac 3 8}g^{\frac 1 8 }},
 \end {equation}
i.e. the scaling obtained by \cite{Bohr_1993} and modified by logarithmic corrections. 

We now consider the regime described in Fig.\ref{fig1}-b that may be obtained in stationary regime with particular superhydrophobic treatment [\cite{Maynes_2011}] or with inverse gravity [\cite{jameson_2010, button_2010}]. In this regime the force opposed to fluid inertia at the boundaries is dictated only by surface tension and not by gravity, there is no developed shock, no liquid "wall". In other words, the flux of momentum is only balanced by surface tension, which means that equations (\ref{e.20}) and (\ref{e.21}) are simply replaced now by:

   \begin{equation}
\label{e.23}
\rho h (r) <u(R_J)^2>\approx \gamma (1-\cos \theta),
 \end {equation}

with $\theta$ the static contact angle

Using eq.(\ref{e.19}) in the limit $r=R_J\gg r_0$, this condition yields a new scaling that reads:

   \begin{equation}
\label{e.24}
R_J=\left(\frac{27\sqrt{3}c^3}{16\pi^6}\right)^{\frac 1 4 } (1-cos \theta)^{\frac 1 4}Q^{\frac 3 4}\nu^{-\frac 1 4}\rho^{\frac 1 4}\gamma^{-\frac 1 4}. 
 \end {equation}

This scaling is the same than the one suggested by \cite{bhagat_2018} and previously by \cite{button_2010}. It explains why the scaling obtained by   \cite{bhagat_2018} applies to the experimental data of  \cite{jameson_2010} even if the theory leading to this scaling is not the right one. 

We thus do not believe that there is a “universal” scaling that should hold for any circular "print" formed around an impacting jet. Sometimes one can find Bohr's scaling and sometimes Baghat and Button one, it is the analysis of the conditions around the impact that will matter.

\section{Another possible occurrence of Bhagat and Button scaling. }
\label{sec.5}

\begin{figure}
\begin{centering}
\includegraphics[width=0.6 \columnwidth]{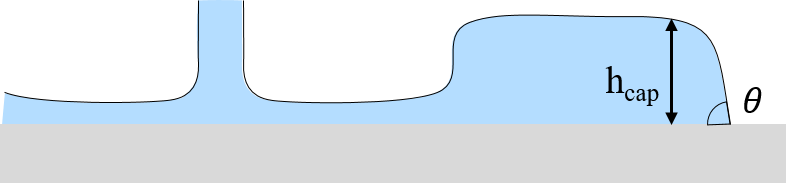}
\caption {sketch of the intermediate regime for a low viscosity liquid in partial wetting. }
\label {fig3}
\end{centering}
\end{figure}

We now show that Bahgat's scaling may also be observed in classical circular hydraulic jumps. In \cite{bhagat_2018} paper, the authors consider an intermediate regime where the liquid has not yet reached the edge of the plate (see Fig.\ref{fig3}). In their experimental evidence the authors consider partial wetting conditions (they use Perspex, glass and Teflon) and aqueous solutions. Given that the front propagation speed is rather small, we can consider that the liquid front height is approximately given by

   \begin{equation}
\label{e.25}
h_{cap}\approx\left(\frac \gamma {\rho g}\right) ^\frac1 2 (1-\cos \theta)^\frac1 2. 
 \end {equation}

Considering the eq.(\ref{e.20}) for low viscosity liquid and moderate flow rate one can conclude that: 
   \begin{equation}
\label{e.26}
H(r)\approx H_\infty\approx h_{cap} .
 \end {equation}

Therefore the (simplified) shock condition (\ref{e.21}) previously obtained leads to: 

  \begin{equation}
\label{e.27}
\rho h (r) <u(R_J)^2>\approx \frac 1 2  \rho g h_{cap}^2.
 \end {equation}

Surprisingly, this argument leads again exactly to the "surface tension dominated" scaling :

   \begin{equation}
\label{e.28}
R_J=\left(\frac{27\sqrt{3}c^3}{16\pi^6}\right)^{\frac 1 4 } (1-cos \theta)^{\frac 1 4}Q^{\frac 3 4}\nu^{-\frac 1 4}\rho^{\frac 1 4}\gamma^{-\frac 1 4}. 
 \end {equation}

Following now a remark from \cite{bhagat_2018}, one can also denote that by defining the Weber number as 

   \begin{equation}
\label{e.29}
\mathrm{We}\approx \frac {\rho h (r) <u(R_J)^2> }\gamma \approx cste, 
 \end {equation}

i.e., a constant Weber number that replaces the constant Froude number encountered in fully established hydraulic jump with a complete, flowing outer film. 

Considering $\theta=\frac \pi 2$, we obtain that: 

   \begin{equation}
\label{e.30}
\mathrm{We} \approx \frac 1 2, 
 \end {equation}
 which is the order of magnitude of the Weber number reported in \cite{bhagat_2018}.

\section{Conclusion}
In summary, we have reconsidered the problem of scaling law selection of the "radius of influence" in the problem of vertical jet impact on a horizontal solid surface. In our opinion, the ideal law (\ref{eq.1}) proposed by Bohr and coworkers (to which, one should not forget to add logarithmic corrections as in \cite{duchesne_2014}) corresponds to the ideal situation of a stationary hydraulic jump formed inside a liquid film extending on the whole solid surface. On the opposite, the scaling (\ref{eq.2}) suggested in ref. [\cite{bhagat_2018}] rather holds in different situations, some of these ones being:

-	stationary impact of a jet on a dry surface, possibly superhydrophobic, without formation of the outer film (atomization ring),

-      stationary impact of a jet on a dry surface in inverse gravity (impact of a jet on a ceiling),

-	transient regime of circular hydraulic jump formation for low viscosity liquids in partial wetting.

It would be interesting to explore in more details these three situations, and to identify possible other ones. In our opinion, there is no need to imagine some universal extra capillary term imposing the scaling (\ref{eq.2}) as imagined in ref [\cite{bhagat_2018}]. Though this extra term really exists, when the control volume contains the free surface of the film instead of excluding it, it is in practice compensated by another one in a way consistent with classical thermodynamics. As usual in free surface flows there is no increase or decrease of velocity that could be due alone to the action of surface tension, except when Marangoni effects are involved [\cite{marmottant_2000}]. Going on in this direction would be just reproducing for thin film flows on a  solid,  the initial mistake of Hagen and Bouasse.  

If we come back to the question of Bohr scaling we have left a bit aside the questions of the logarithmic corrections and the possible existence of a critical Froude number at the jump exit, suggested in [\cite{duchesne_2014}]. The possible existence of this critical Froude number leads to a different exponent for the Logarithmic corrections ($3/8$ instead of $1/8$) and this question is still not solved. As told in the introduction, recent non-linear analytical treatment of the film flow suggests that such a critical Froude number could exist, but this remains to be established and convincingly explained. 

A specific problem of great interest where these considerations should mater is the question of jet impacts on inclined plates. It is not obvious in this kind of problem that a perfect hydraulic jump can exist, or not, and the two scaling should compete against each other in a  way that merits to be investigated. The influence of a external fields, here the tangent component of gravity on a circular shock is a fundamental question of great interest.  A specific effort should be done in this direction [\cite{wilson_2012, duchesne_2013_ij}]. 

Declaration of Interests. The authors report no conflict of interest.

\bibliographystyle{jfm}
\bibliography{bibliojump}
\end{document}